\documentclass[10pt,preprintnumbers,aps,amssymb,nofootinbib,amsmath,superscriptaddress,notitlepage,prd]
{revtex4-2}
\usepackage{epsfig,epsf}
\usepackage{comment}
\usepackage{bm} 
\usepackage{color} 
\usepackage{slashed}
\usepackage{relsize}	
\usepackage{soul} 
\usepackage{hyperref}
\usepackage{tensor} 
\usepackage{yfonts} 
\usepackage{enumitem} 
\newcommand{\beq}{\begin{equation}}
\newcommand{\beql}[1]{\begin{equation}\label{#1}}
\newcommand{\eeq}{\end{equation}}
\def\bal#1\gal{\begin{align}#1\end{align}}
\newcommand{\ball}[1]{\bal\label{#1}}
\newcommand{\eq}[1]{(\ref{#1})}
\newcommand{\fig}[1]{Fig.~\ref{#1}}
\renewcommand{\sec}[1]{Sec.~\ref{#1}}
\DeclareMathOperator{\sgn}{sgn}

\DeclareMathOperator{\Tr}{\mathrm{Tr}}

\renewcommand{\b}[1]{{\bm #1}} 
\newcommand{\unit}[1]{\hat {{\bm #1}}} 

\begin{document}

\title{Quark and gluon production in the presence of the time-varying chiral magnetic current}

\author{Kirill Tuchin}

\affiliation{
Department of Physics and Astronomy, Iowa State University, Ames, Iowa, 50011, USA}

\date{\today}


\begin{abstract}

The chiral magnetic effect (CME) consists in the induction of the electric current along the direction of the magnetic field.  The corresponding transport coefficient $b_0$, known as the chiral magnetic conductivity, is proportional to the chiral imbalance in the medium. In many systems, such as quark-gluon plasma, $b_0$ is time-dependent. This paper studies the effect of the time variation of $b_0$ on the particle spectra and energy loss produced through the chiral Cherenkov and associated processes in Abelian and non-Abelian systems. The rates of all processes are derived in the ultrarelativistic approximation. The results are applied to the relativistic heavy-ion collisions utilizing a specific model describing the relaxation of the initial $P$-odd domain within the quark-gluon plasma. The corresponding energy loss is computed. The results suggest strong polarization of jets in quark-gluon plasma.

\end{abstract}

\maketitle

\section{Introduction}\label{sec:a}

An ultrarelativistic particle losses most of its energy in a medium due to radiation. The amount of energy lost and the spectrum  of the emitted radiation reflect the medium properties. This paper investigates novel mechanisms of energy loss in chiral medium, which contains chiral fermions. The peculiar properties  of energy loss in a chiral medium are due to the chiral anomaly, which induces P and CP-odd transport phenomena \cite{Kharzeev:2013ffa}. One such phenomenon is the chiral magnetic effect (CME), whereby electric current is induced in the direction of an external magnetic field: $\b j= b_0 \b B$ \cite{Kharzeev:2004ey,Kharzeev:2007jp,Kharzeev:2009fn,Kharzeev:2007tn,Fukushima:2008xe,Barredo-Alamilla:2023xdt,vonDossow:2025fbb}. Another closely related phenomenon is the anomalous Hall effect, where a current is induced in the direction perpendicular to an external electric field: $\b j = \b b\times \b E$. The chiral magnetic conductivity $b_0$ is proportional to the chiral chemical potential $\mu_5$, which reflects the net chirality in the medium. The non-Abelian version of the chiral magnetic effect was investigated in \cite{Akamatsu:2013pjd,Duari:2025kar}.

As a result of the chiral magnetic effect, new channels of energy loss are opened in empty space due to the momentum supplied by the chiral magnetic current.  These processes include the chiral Cherenkov radiation and its cross channels, which have been studied in Ref.~\cite{Tuchin:2018sqe,Huang:2018hgk,Tuchin:2018mte,Hansen:2020irw,Hansen:2022nbs,Hansen:2023wzp,Hansen:2024kvc}, and recently reviewed in \cite{Hansen:2024xdg}. The cascade induced by these novel channels was studied in \cite{Hansen:2025gzt}.  Additionally, the non-Abelian version of the chiral Cherenkov radiation was explored in \cite{Hansen:2024rlj}. While the anomalous Hall current also contributes to energy loss, as discussed in \cite{Huang:2018hgk,Tuchin:2018mte,Hansen:2024kvc,Tuchin:2025stl}, the present paper assumes a spatially uniform medium, so we disregard the anomalous Hall currents in our analysis.

The references mentioned above ignore the time-variation of the chiral magnetic conductivity $b_0$. However, in the quark-gluon plasma produced in relativistic heavy-ion collisions the chirality is sourced by the parallel chromoelectric and chromomagnetic fields in the initial stage following a heavy-ion collision \cite{Lappi:2017skr} or by sphaleron transitions during the plasma's evolution \cite{Fukushima:2008xe,Arnold:1996dy,Arnold:1998cy,Bodeker:1998hm}. These processes introduce the time-dependence of $b_0$.  The time-dependence of $b_0$ can also be driven by the chiral plasma instability \cite{Joyce:1997uy,Boyarsky:2011uy,Hirono:2015rla,Akamatsu:2013pjd,Tuchin:2014iua,Buividovich:2015jfa,Manuel:2015zpa,Kirilin:2017tdh}. This motivated my recent study of photon production in the presence of time-dependent chiral imbalance \cite{Tuchin:2025stl,Tuchin:2025bll}. This paper builds upon that work and presents a comprehensive calculation of all processes contributing to energy loss in both Abelian and non-Abelian media. I derive general expressions for the rates of all parton splitting processes at any $b_0(t)$ and then apply these expressions to a specific model describing particle production in the presence of a decaying $P$-odd domain in heavy-ion collisions. 

The paper is organized as follows. In \sec{sec:b}, the photon and gluon wave functions in the presence of the chiral magnetic current are derived. \sec{sec:c} computes the rates of the various splitting channels. Fig.~\fig{fig:qqg}, \fig{fig:gqq}, and \fig{fig:ggg} exhibit the gluon and quark spectra produced in the processes $q\to q+g$, $g\to q+\bar q$ and $g\to g+g$ respectively. The energy loss induced by the chiral magnetic current is computed in \sec{sec:m} and is shown in \fig{fig:eloss}. 
The concluding remarks are presented in \sec{sec:s}.

I employ the natural units $\hbar=c=1$, $\alpha=g^2/4\pi$, and the notation $p^\mu = (E, \b p)$, where $p\equiv |\b p|$.

\section{Photon and gluon wave functions in the presence of CME}\label{sec:b}

\subsection{QED}\label{ba}

Perturbative calculations  in chiral medium in the presence of the chiral magnetic current are facilitated  by an effective field theory obtained by adding the Chern-Simons term to the Maxwell Lagrangian \cite{Wilczek:1987mv,Carroll:1989vb, Sikivie:1984yz}:
\ball{ba1}
\mathcal{L}= -\frac{1}{4} \left(F_{\mu\nu}^2+c_1\theta F_{\mu\nu} \tilde F^{\mu\nu}\right)\,,
\gal
where $F_{\mu\nu}$ is the electromagnetic field tensor, $\tilde F^{\mu\nu}= \frac{1}{2}\epsilon_{\mu\nu\lambda\rho} F^{\lambda\rho}$ is its dual, and $c_1$ is the chiral anomaly coefficient given by $c_1=\frac{e^2}{2\pi^2}N_c\sum_f q_f^2$. The Chern-Simons term is coupled to an external pseudoscalar field $\theta$ chosen to reproduce the chiral magnetic and anomalous Hall currents. Namely, $c_1\partial_\mu\theta = b_\mu=(b_0,-\b b)$, where  $b_\mu$ is a function of time but not  spatial coordinates. The axial chemical potential $\mu_5$ is identified with $\dot \theta$, so that $b_0=c_1\mu_5$. 
The medium is assumed to be isotropic, implying that $\b b=0$. Perturbation theory in the presence of finite $\b b$ has been discussed in \cite{Huang:2018hgk,Tuchin:2018mte,Hansen:2024kvc,Tuchin:2025stl}. 

The electromagnetic field, in the radiation gauge $A^0=0$, $\b \nabla\cdot \b A=0$, is described by the vector potential $\b A$ that  satisfies the equation:
\ball{ba2}
\nabla^2 \b A -\partial_t^2\b A+ b_0(t)\b \nabla\times \b A=0\,.
\gal
It is assumed that the chiral magnetic conductivity $b_0$ is a given function of time that is unaffected by the radiation. The wave function of a photon of energy $\omega$, momentum $\b k$ and polarization $\lambda$ is a plane wave solution to \eq{ba2}: 
\ball{ba3}
\b A_{\b k\lambda}(t, \b r)= \frac{1}{\sqrt{V}}a(t)\b \epsilon_\lambda e^{ i\b k\cdot \b r}\,.
\gal
The photons must be circularly polarized with the polarization vector $\b\epsilon_\lambda$, $\lambda=\pm 1$, which obeys the equation $i\unit k \times \b\epsilon_\lambda = \lambda \b\epsilon_\lambda$. The product of the photon polarization vectors satisfies the following identity:
\bal\label{bb5}
\b\epsilon^{i*}_\lambda \b\epsilon^{j}_\lambda = 
\frac{1}{2}\left[\left( \delta^{ij}-\frac{k^ik^j}{k^2}\right) +i\lambda\varepsilon^{ij\ell}\frac{k^\ell}{ k}\right]\,,
\gal
where no summation over $\lambda$ is implied.

 Substituting \eq{ba3} into \eq{ba1} one finds that the amplitude $a(t)$ is governed by the equation
\ball{ba5}
\ddot a(t)+\Omega^2(t)a(t) =0\,,
\gal
where 
\ball{ba7}
\Omega^2(t)= k^2 -\lambda b_0(t)k\,.
\gal

Eq.~\eq{ba5} can be solved in a general form, assuming that $\Omega(t)$ is slowly varying function (see e.g.\ \cite{Tuchin:2025stl}): 
\bal\label{ba9} 
a(t)= \frac{1}{\sqrt{2\Omega}}e^{- i\int_0^t \Omega(t')dt'}\,.
\gal
The amplitude $a$ is normalized such that the wave function \eq{ba3} represents one photon per unit volume.  Expanding $\Omega$ in the ultrarelativistic limit $k\gg b_0$:
\bal\label{ba11}
\Omega(t)\approx k-\frac{\lambda b_0(t)}{2}
\gal
leads to the final result for the amplitude:
\bal\label{ba13}
a(t)= \frac{1}{\sqrt{2k}}e^{- ikt +\frac{i\lambda}{2}\int_0^t b_0(t')dt'}\,.
\gal 
The validity of \eq{ba13} is restricted to photon momenta $k$ that satisfy the condition:
\ball{ba15}
\left| \dot b_0\right|\ll k^2\,,
\gal
where the dot indicates the time-derivative.

The scattering matrix elements are proportional to the Fourier image of the amplitude: 
\ball{ba17}
\tilde a(q)=  \int_{-\infty}^{+\infty} e^{i(q+i\epsilon\sgn t)t}a(t)dt\,.
\gal
The positive small parameter $\epsilon$ ensures convergence of the integral. It indicates the finite width of the quasistationary energy levels at constant $b_0$ and estimated to be of the order $g^2 b_0$\cite{Hansen:2023wzp}. Substituting \eq{ba13} into \eq{ba17} one obtains:
\ball{ba19}
\tilde a(q)=\frac{2}{\sqrt{2k}}\int_0^\infty e^{-\epsilon t+ \frac{i\lambda}{2}\int_0^t b_0''(t')dt'}\cos\left[ (k-q)t-\frac{\lambda}{2}\int_0^t b_0'(t')dt'\right]dt\,,
\gal
where the functions
\bal
b_0'(t) &= \frac{b_0(t)+b_0(-t)}{2}\,,\label{ba20}\\
b_0''(t)&= \frac{b_0(t)-b_0(-t)}{2}\,,\label{ba21}
\gal
are respectively the even and odd components of $b_0(t)$.

The wave functions of fermions are not modified and given by the standard expressions. Hence the wave function of an incident fermion with momentum $\b p$ and spin $s$ reads
\ball{ba29}
\psi_{\b ps} (t, \b r)= \frac{1}{\sqrt{2EV}}u_s(\b p)e^{-iEt+i \b p\cdot \b r}\,,
\gal
where $E=\sqrt{p^2+m^2}$.

\subsection{QCD}\label{bb}

The color version of the chiral magnetic effect, which describes the induction of the color current in the direction of the color magnetic field, can be similarly described by adding the corresponding non-Abelian Chern-Simons term to the Yang-Mills Lagrangian. The effective Lagrangian is given by:
\ball{bb1}
\mathcal{L}= -\frac{1}{2} \Tr\left(F_{\mu\nu} F^{\mu\nu}\right)-\frac{c_2}{2} \theta \Tr\left(  F_{\mu\nu}\tilde F^{\mu\nu}\right)  \,,
\gal
The field tensor and the potentials are defined as usual:
\ball{bb2}
F_{\mu\nu}= \partial_\mu A_\nu -\partial_\nu A_\mu-ig[A_\mu, A_\nu]\,,
\gal
$A_\mu = A_\mu^at^a$, $F_{\mu\nu}= F_{\mu\nu}^at^a$, where $t^a$ are the SU(3) generators and the anomaly coefficient is $c_2=\frac{g^2N_f}{4\pi^2}$. 

The linearized equations of motion clearly have the same form as those discussed in the preceding section for the Abelian case. The corresponding plane wave solutions are given by
\ball{bb4}
\b A^b_{\b k\lambda}(t, \b r)= \frac{1}{\sqrt{V}}a(t)e^b\b \epsilon_\lambda e^{ i\b k\cdot \b r}\,.
\gal
Here, $e^b$ is a unit vector in the adjoint color representation, which we will omit from the equations henceforth, as per usual convention. The amplitude $a(t)$ is given by \eq{ba13} in the ultrarelativistic approximation.

Unlike the Abelian Chern-Simons term, the non-Abelian one includes a cubic term that contributes to the triple-gluon interaction vertex. Substituting the identity
\ball{bb6}
\Tr\left( F^{\mu\nu}\tilde F^{\mu\nu}\right)= 2 \epsilon^{\mu\nu\rho\sigma}\partial_\mu\Tr\left(A_\nu\partial_\rho A_\sigma-g\frac{2i}{3}A_\nu A_\rho A_\sigma\right) 
\gal 
into the second term in \eq{bb1} and integrating by parts yields 
\ball{bb8}
\mathcal{L}_\theta= c_2\partial_\mu\theta \epsilon^{\mu\nu\rho\sigma}\left(\frac{1}{2}A^a_\nu\partial_\rho A^a_\sigma-g\frac{2i}{3}\frac{1}{4}if^{abc}A^a_\nu A^b_\rho A^c_\sigma\right)\,.
\gal
The first term in \eq{bb6} contributes to the equation of motion, while the second one to the triple-gluon vertex. The corresponding  Feynman rule for the anomalous contribution to the triple-gluon vertex in momentum space is:
\ball{bb10}
\text{anomalous triple-gluon vertex}=gb_\mu\epsilon^{\mu\nu\rho\sigma}f^{abc}\,,
\gal
where we follow the conventions of  \cite{Peskin:1995ev}. It was shown in \cite{Hansen:2024rlj} that this contribution is suppressed in the ultrarelativistic limit by the inverse power of particles' energy. 

In summary, in the ultrarelativistic limit the effect of anomaly is to impart a time- and helicity-dependent phase 
\ball{bb13}
e^{\frac{i\lambda}{2}\int_0^t b_0(t')dt'} 
\gal
to the photon and gluon fields, as compared to their expressions in empty space. Note also that all medium effects other than the anomalous chiral magnetic and Hall currents are omitted for simplicity. 

\subsection{Plasma oscillations}

Thus far, we ignored the collective medium effects on the wave functions of gauge bosons. In the ultrarelativistic limit discussed in this work, these effects can be described by the plasma frequency $\omega_p$. In the quark-gluon plasma at temperature $T$ it is given by $\omega_p^2=\frac{g^2T^2}{18}(2N_c+N_f)$ \cite{RISCHKE2004197}.

The plasma oscillations modify Eq.~\eq{ba2}, which is obeyed by the electromagnetic vector potential, as follows: 
\ball{bc1}
\nabla^2 \b A -\partial_t^2\b A+ b_0(t)\b \nabla\times \b A - \omega_p^2\b A=0\,.
\gal
As a result, the function $\Omega(t)$ reads
\ball{bc3}
\Omega(t)= \sqrt{k^2-\lambda b_0(t)k+\omega_p^2}\approx k-\frac{\lambda b_0}{2}+\frac{\omega_p^2}{2k}\,,
\gal
where we assumed that $k\gg \omega_p\gg |b_0|$ as favored by the phenomenological applications. A similar equation holds for the gluon field. In place of \eq{ba13}, we obtain for the amplitude:
\bal\label{bc5}
a(t)= \frac{1}{\sqrt{2k}}e^{- ikt +\frac{i\lambda}{2}\int_0^t b_0(t')dt' - \frac{i\omega_p^2t}{2k}}\,.
\gal 

To keep equations less bulky, the plasma frequency will be omitted in all expressions unless otherwise indicated.

\section{Parton spectra}\label{sec:c}

Now, I move on to the main section of the paper, where the expressions for the photon and gluon fields derived in the previous section are utilized to calculate the differential probabilities of photon, gluon, and quark production.

\subsection{$q\to q+\gamma$ and $q\to q+g$}\label{sec:ca}

First, consider photon radiation by an ultrarelativistic quark. The initial and final quark momenta are $p_z\unit z$ and $\b p’$, respectively. The photon's momentum is $\b k = x p_z \unit z+ \b k_\bot$, where $x$ is the longitudinal momentum fraction it carries and $\b k_\bot$ its component in the $xy$ plane. The corresponding energies are
\bal
E&= p_z\left( 1+ \frac{m^2}{2p_z^2}\right)\,,\label{ca1}\\
E'&= (1-x)p_z\left( 1+\frac{m^2+k_\bot^2}{2(1-x)^2p_z^2}\right)\,,\label{ca2}
\gal 
where the terms of the order $k_\bot^2/p_z^2$ and higher are neglected. The energy is not conserved unless $b_0$ is constant. 

The  element of the scattering matrix describing photon radiation reads
\ball{ca5}
S(q\to q\gamma)&= e \int \bar \psi_{\b p's'} \b \gamma\cdot \b A^*_{\b k\lambda}\psi_{\b ps} d^3x dt
=e\frac{\bar u _{s'}(\b p')\b\gamma\cdot \b \epsilon^*_\lambda u_s(\b p)}{2V^{3/2}\sqrt{EE'}}(2\pi)^3\delta(\b p-\b p'-\b k)\tilde a^*(E-E')\,,
\gal
where $\tilde a$ is given by \eq{ba19} and $e$ is the quark's electric charge. The photon radiation probability is then calculated as
\ball{ca8}
dw &= \frac{1}{2}\sum_{\lambda,s,s'}|S(q\to q\gamma)|^2\frac{V d^3p'}{(2\pi)^3}\frac{V d^3k}{(2\pi)^3}= \frac{e^2}{8(2\pi)^3 E}\sum_\lambda |\tilde a(E-E')|^2d^2k_\bot \frac{dx}{1-x} \Tr[(\slashed p+m)\slashed \epsilon^*_\lambda (\slashed p'+m)\slashed \epsilon_\lambda]\,,
\gal
which, upon using  \eq{ca1},\eq{ca2} and \eq{bb5}, yields the differential probability \cite{Tuchin:2025stl}:
\ball{ca9}
\frac{dw(q\to q\gamma)}{d^2k_\bot dx}&= \frac{1}{8(2\pi)^3 E}\sum_\lambda |\tilde a(E-E')|^2\frac{2e^2}{x(1-x)^2}\left\{ k_\bot^2\frac{1+(1-x)^2}{x}+m^2 x^3\right\}\,. 
\gal
Using \eq{ba19} and taking into account that 
\ball{ca10}
k-E+E'= \frac{k_\bot^2+x^2m^2}{2x(1-x)p_z}\,,
\gal
one obtains:
\ball{ca11}
\tilde a(E-E')=\frac{2}{\sqrt{2k}}\int_0^\infty e^{-\epsilon t+ \frac{i\lambda}{2}\int_0^t b_0''(t')dt'}\cos\left[ \frac{k_\bot^2+x^2m^2}{2x(1-x)p_z}t-\frac{\lambda}{2}\int_0^t b_0'(t')dt'\right]dt\,.
\gal

The gluon spectrum emitted by an ultrarelativistic quark differs from \eq{ca9} only by a color factor and the replacement $e\to g$:
\ball{ca19}
\frac{dw(q\to qg)}{d^2k_\bot dx}&= C_F\frac{dw(q\to q\gamma)}{d^2k_\bot dx}\bigg|_{e\to g}\,,
\gal
where $x$ and $k_\bot$ represent the longitudinal momentum fraction and the transverse momentum of the appropriate gauge boson.

\subsubsection{Constant $b_0$}

At a constant $b_0$,  $\tilde a(q)$ is proportional to the delta function expressing energy conservation. After some algebra,  \eq{ca9} reduces to \cite{Tuchin:2025stl}:
\ball{ca13}
\frac{d\dot w(q\to q\gamma)}{dx d^2k_\bot}= \frac{e^2}{16\pi^2 E } \sum_\lambda\left\{ \lambda E b_0 \frac{1+(1-x)^2}{x} -2m^2\right\}
\delta\left( k_\bot^2-\lambda b_0 E x(1-x)+m^2x^2\right)\,.
\gal
The first term in the braces of \eq{ca9} and \eq{ca13} is proportional to the splitting function $P_{\gamma q}(x)$. 

\subsubsection{$b_0(t)=A+ B\tanh\frac{t}{\tau}$}

In this specific model, $A$ and $B$ are real constants and $\tau$ is a positive constant. The sign of $B$ determines whether $b_0$ increases or decreases over time, starting from its initial value $A-B$ and ending at its final value $A+B$. In particular, this model describes the decay of the initial chirality when $A=-B>0$ or the generation of a chiral domain in the initially chirality-neutral medium when $A=B>0$.

Given the function $b_0(t)$ specified in this subsection's title, the integral in the amplitude \eq{ca11} have this general form:
\ball{j3}
I=\int_0^\infty  e^{-\epsilon \tau}e^{\frac{i\Lambda }{2}B\tau \log\cosh\frac{t}{\tau}}
\cos\left( \alpha t - \frac{\Lambda}{2}A t \right) dt\,.
\gal
The parameters $\Lambda$ and $\alpha$ vary depending on the specific reaction. This integral  can be performed using Eq.~1.9(5) of \cite{bateman1954tables} and yields the result:
\ball{j5}
I=\frac{\tau 2^{-i\Lambda B \tau/2-2}}{\Gamma(-i\Lambda B \tau/2)}
\Gamma\left[\frac{1}{2}\left(-\frac{i\Lambda \tau}{2}(A+B)+i\alpha \tau+\epsilon\tau\right) \right]
\Gamma\left[\frac{1}{2}\left(\frac{i\Lambda \tau}{2}(A-B)-i\alpha \tau+\epsilon\tau\right) \right]\,.
\gal
In particular, replacing  $\Lambda= \lambda$, $\alpha=k-E+E'$ and using \eq{ca10} yields for the amplitude \eq{ca11}:
\ball{caj1}
\tilde a(E-E')= \frac{\tau}{\sqrt{2k_z}}\frac{ 2^{-i\lambda B \tau/2-1}}{\Gamma(-i\lambda B \tau/2)}
\Gamma\left[\frac{i\tau}{2}\left(-\frac{\lambda }{2}(A+B)+\frac{k_\bot^2+x^2m^2}{2x(1-x)p_z} -i\epsilon\right) \right]\nonumber\\
\times
\Gamma\left[\frac{i\tau}{2}\left(\frac{\lambda }{2}(A-B)-\frac{k_\bot^2+x^2m^2}{2x(1-x)p_z} -i\epsilon\right) \right]\,.
\gal
The final formula for the photon spectrum is:
\ball{caj3}
\frac{dw(q\to q\gamma)}{d^2k_\bot dx}&= \frac{1}{8(2\pi)^3 E}\sum_\lambda \frac{2e^2}{x(1-x)^2}\left\{ k_\bot^2\frac{1+(1-x)^2}{x}+m^2 x^3\right\} \frac{\tau^2}{2k_z}\frac{ 1}{4\left|\Gamma(-i\lambda B \tau/2)\right|^2}\nonumber\\
&\times\left|\Gamma\left[\frac{i\tau}{2}\left(-\frac{\lambda }{2}(A+B)+\frac{k_\bot^2+x^2m^2}{2x(1-x)p_z} -i\epsilon\right) \right]\Gamma\left[\frac{i\tau}{2}\left(\frac{\lambda }{2}(A-B)-\frac{k_\bot^2+x^2m^2}{2x(1-x)p_z} -i\epsilon\right) \right]\right|^2\,.
\gal

\begin{figure}[ht]
\begin{tabular}{cc}
      \includegraphics[width=0.45\linewidth]{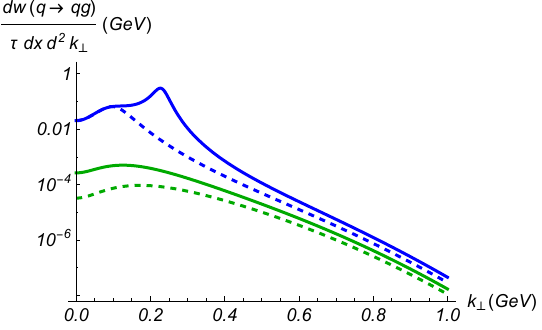} &
       \includegraphics[width=0.45\linewidth]{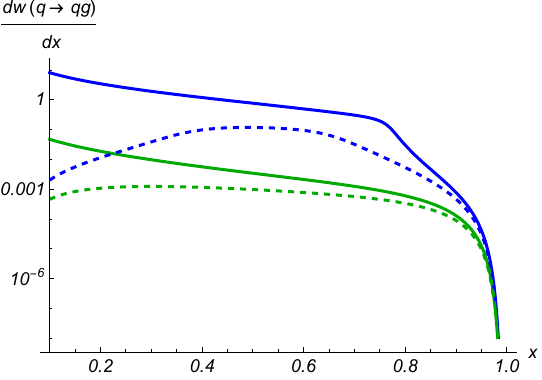} \
      \end{tabular}
  \caption{The gluon spectrum produced by the process $q\to q+g$. The blue lines represent $\lambda=1$, the green line  $\lambda=-1$, the solid lines indicate  $\omega_p=0$, and the dashed line  $\omega_p=0.3$~GeV. The chiral magnetic conductivity is $b_0(t)=A_1+ B_1\tanh\frac{t}{\tau}$ with $A=10$~MeV, $B=-5$~MeV, and $\tau=25$~GeV$^{-1}\approx$ 5~fm/c.  The resonance width $\epsilon=1$~MeV. The incident quark's energy is $E=20$~GeV and mass $m=0.3$~GeV. In the left panel, it is assumed that the gluon carries the fraction $x=0.5$ of the incident quark’s longitudinal momentum.   
  }\label{fig:qqg}
\end{figure}

The gluon spectrum, obtained using $\eq{ca19}$,  is depicted in \fig{fig:qqg}, with the color indicating gluon polarization. 


The effect of plasma oscillations is described by the last term in the right-hand-side of \eq{bc3}.  This term appears in \eq{ca10} and effectively shifts the transverse momentum  $k^2_\bot \to k^2_\bot + (1-x)\omega_p^2$ in \eq{ca11}. This replacement must be made in the arguments of the gamma functions in \eq{caj3}.  The corresponding gluon spectrum is shown by the dashed lines in \fig{fig:qqg}. The finite $\omega_p$ reduces the total gluon multiplicity because it shrinks the phase space available for the chiral Cherenkov effect. Nevertheless, it does not diminish its strong polarization.


\subsection{$\gamma\to q+\bar q$ and $g\to q+\bar q$}\label{sec:cb}

The quark and antiquark pair production process is the cross channel of photon production. Given the incident photon's momentum  $\b k$ and the momenta of the quark and antiquark as $\b p$ and $\b p’$ respectively, the corresponding $S$-matrix element can be expressed as:
\ball{cb1}
S(\gamma\to q\bar q)&= e \int \bar \psi_{\b p's'} \b \gamma\cdot \b A_{\b k\lambda}\psi_{\b ps} d^3x dt
=e\frac{\bar u _{s}(\b p)\b\gamma\cdot \b \epsilon_\lambda v_{s'}(\b p')}{2V^{3/2}\sqrt{EE'}}(2\pi)^3\delta(\b p+\b p'-\b k)\tilde a(E+E')\,,
\gal
In the context of pair production, the longitudinal momentum fraction is defined as $x = p_z/k_z$, where the incident photon’s momentum is assumed to be in the $z$ direction. 

The pair-production probability reads
\bal
dw(\gamma\to q\bar q) &= \frac{1}{2}\sum_{\lambda,s,s'}|S(\gamma\to q\bar q)|^2\frac{V d^3p'}{(2\pi)^3}\frac{V d^3p}{(2\pi)^3}= \frac{e^2}{8(2\pi)^3}\sum_\lambda |\tilde a(E+E')|^2 \frac{d^3p}{x(1-x)k_z^2} \Tr[(\slashed p+m)\slashed \epsilon^*_\lambda (\slashed p'-m)\slashed \epsilon_\lambda]\,.\label{cb3}
\gal
Using \eq{bb5} and  the expressions for $E$ and $E'$ given by 
\bal
E&= xk_z\left( 1+ \frac{m^2+p_\bot^2}{2x^2k_z^2}\right)\,,\label{cb6}\\
E'&= (1-x)k_z\left( 1+\frac{m^2+p_\bot^2}{2(1-x)^2k_z^2}\right)\,,\label{cb7}
\gal 
we obtain the differential probability of quark production via photon decay:
\bal
\frac{dw(\gamma\to q\bar q)}{d^2p_\bot dx}&= \frac{1}{8(2\pi)^3 k_z}\sum_\lambda |\tilde a(E+E')|^2\frac{2e^2}{x^2(1-x)^2}\left\{ p_\bot^2\left[x^2+(1-x)^2\right]+m^2\right\}\,. \label{cb9}
\gal
Since 
\ball{cb9.5}
k-E-E'= -\frac{p_\bot^2+m^2}{2x(1-x)k_z}\,,
\gal
\eq{ba19} implies:
\ball{cb10}
\tilde a(E+E')=\frac{2}{\sqrt{2k}}\int_0^\infty e^{-\epsilon t+ \frac{i\lambda}{2}\int_0^t b_0''(t')dt'}\cos\left[- \frac{p_\bot^2+m^2}{2x(1-x)k_z}t-\frac{\lambda}{2}\int_0^t b_0'(t')dt'\right]dt\,.
\gal

The spectrum of quarks produced through gluon decay reads:

\ball{cb21}
\frac{dw(g\to q\bar q)}{d^2p_\bot dx}= \frac{1}{2}\frac{dw(\gamma\to q\bar q)}{d^2p_\bot dx}\bigg|_{e\to g}\,,
\gal
where 1/2 is the color factor.

\subsubsection{Constant $b_0$}

At constant $b_0$, the amplitude \eq{bc5} reduces to $a(t)= e^{-i\omega t}/\sqrt{2k_z}$, where $\omega= k_z-\lambda b_0/2$. As the result, $\tilde a(E+E')= 2\pi\delta(\omega-E-E')/\sqrt{2k_z}$. The square of this quantity involves the product of two identical delta functions. One of them is interpreted as $(2\pi)^{-1}$ times the observation time, whereas the other one can be cast in the following form:
\ball{cb11}
\delta(\omega-E-E')= 2k_zx(1-x)\delta\left(m^2+p_\bot^2+\lambda k_z b_0x(1-x)\right)\,.
\gal
Using these results in \eq{cb9} yields the quark production rate at constant $b_0$:
\ball{cb14}
\frac{d\dot w(\gamma\to q\bar q)}{dx d^2k_\bot}= \frac{e^2}{16\pi^2 k_z x(1-x)} \sum_\lambda\left\{ p_\bot^2 \left[ x^2+(1-x)^2\right]+m^2\right\}
\delta\left( p_\bot^2-\lambda b_0 k_z x(1-x)-m^2\right)\,.
\gal
Eq.~\eq{cb14} agrees with the result obtained in \cite{Tuchin:2018sqe}. Notably, the expression in the square brackets is proportional to the splitting function $P_{qg}(x)$.

\subsubsection{$b_0(t)=A+ B\tanh\frac{t}{\tau}$}

\begin{figure}[ht]
\begin{tabular}{cc}
      \includegraphics[width=0.45\linewidth]{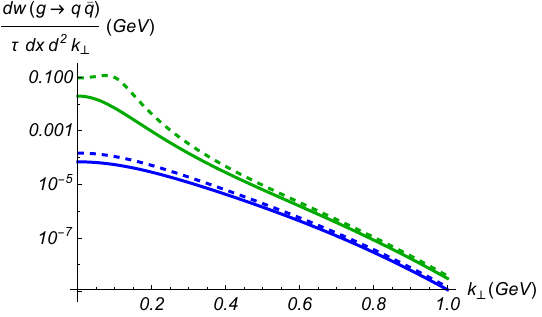} &
       \includegraphics[width=0.45\linewidth]{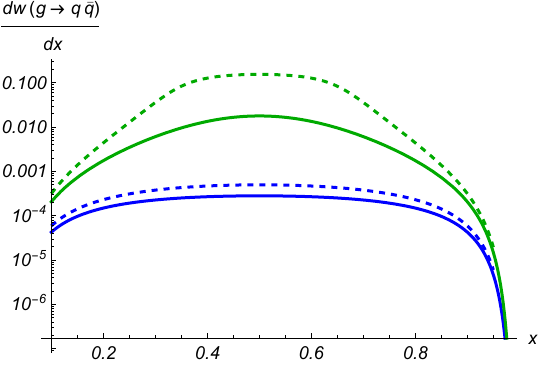} \
      \end{tabular}
  \caption{The quark spectrum produced by the process $g\to q+\bar q$.  The blue lines represent $\lambda=1$, the green lines  $\lambda=-1$, the solid lines indicate  $\omega_p=0$, and the dashed lines  $\omega_p=0.3$~GeV.  The chiral magnetic conductivity $b_0(t)=A_1+ B_1\tanh\frac{t}{\tau}$ with $A=10$~MeV, $B=-5$~MeV, and $\tau=25$~GeV$^{-1}\approx$ 5~fm/c.  The resonance width $\epsilon=1$~MeV. The incident gluon's energy is $E=20$~GeV. The quark's mass is $m=0.3$~GeV and it carries $x=0.5$ of the incident gluon's longitudinal momentum.}\label{fig:gqq}
\end{figure}

In this case, the integral in \eq{cb10} can be evaluated using \eq{j5} with $\Lambda=\lambda$, $\alpha=k-E-E'$, where the latter is given by \eq{cb9.5}. One obtains:
\ball{ja11}
\tilde a(E+E')= \frac{\tau}{\sqrt{2k_z}}\frac{ 2^{-i\lambda B \tau/2-1}}{\Gamma(-i\lambda B \tau/2)}
\Gamma\left[\frac{i\tau}{2}\left(-\frac{\lambda }{2}(A+B)-\frac{p_\bot^2+m^2}{2x(1-x)k_z} -i\epsilon\right) \right]\nonumber\\
\times
\Gamma\left[\frac{i\tau}{2}\left(\frac{\lambda }{2}(A-B)+\frac{p_\bot^2+m^2}{2x(1-x)k_z} -i\epsilon\right) \right]\,.
\gal
Substituting \eq{ja11} into \eq{cb9} yields the quark spectrum in photon decay:
\ball{cbj1}
\frac{dw(\gamma\to q\bar q)}{d^2p_\bot dx}&= \frac{1}{8(2\pi)^3 k_z}\sum_\lambda\frac{2e^2}{x^2(1-x)^2}\left\{ p_\bot^2\left[x^2+(1-x)^2\right]+m^2\right\}\frac{\tau^2}{2k_z}\frac{ 1}{4\left|\Gamma(-i\lambda B \tau/2)\right|^2}\nonumber\\
&
\times \left|\Gamma\left[\frac{i\tau}{2}\left(-\frac{\lambda }{2}(A+B)-\frac{p_\bot^2+m^2}{2x(1-x)k_z} -i\epsilon\right) \right]
\Gamma\left[\frac{i\tau}{2}\left(\frac{\lambda }{2}(A-B)+\frac{p_\bot^2+m^2}{2x(1-x)k_z} -i\epsilon\right) \right]\right|^2\,.
\gal
The quark spectrum in gluon decay is computed according to \eq{cb21} and is represented by solid lines in \fig{fig:gqq}. 

The effect of plasma oscillations on pair production is taken into account by replacing $p_\bot^2\to p_\bot^2-\omega_p^2x(1-x)$. The resulting spectrum is exhibited as dashed lines in \fig{fig:gqq}. One observes that the quark multiplicity has increased compared with the $\omega_p=0$ case. This is because $\omega_p$ acts as a photon mass, thereby expanding the phase space available for pair production.


\subsection{$g\to g+g$}\label{sec:dc}

The initial gluon carries momentum $\b p$ and polarization $\lambda_0$, whereas the two final gluons have momenta $\b k$, $\b p’$, and their respective polarizations are $\lambda$, $\lambda’$. The longitudinal momentum fraction is defined as $x= k_z/p_z$. It is convenient to choose the reference frame such that 
\ball{dc1}
\b p &= (0,0,p_z)\,,\\
\b k&= (k_\bot,0,k_z)\,,\\
\b p'&= (-k_\bot,0,p'_z)\,.
\gal
The corresponding polarization vectors read:
\bal
\b \epsilon_{\lambda_0}(p)&=\frac{1}{\sqrt{2}}(1,\lambda_0i,0)\,,\label{dc3a}\\
\b \epsilon_{\lambda}(k)&=\frac{1}{\sqrt{2}}\left(1,\lambda i,-\frac{k_\bot}{k_z}\right)\,,\label{dc3b}\\
\b \epsilon_{\lambda'}(p')&=\frac{1}{\sqrt{2}}\left(1,\lambda' i,\frac{k_\bot}{p_z'}\right)\,. \label{dc3c}
\gal

The $S$-matrix element describing gluon production in the three-gluon process is given by: 
\ball{dc5}
S(g\to gg)= -g \int f^{abc}
\left\{ g^{\mu\nu}(\partial^\rho A^{a}_{\mu}) A^{b*}_\nu A^{c*}_\rho - g^{\mu\nu} A^{a}_\mu (\partial^\rho A^{b*}_\nu) A^{c*}_\rho\right. \nonumber\\
+
g^{\nu\rho}(\partial^\mu A^{b*}_\nu) A^{c*}_\rho A^{a}_\mu - g^{\nu\rho} A^{b*}_\nu (\partial^\mu A^{c*}_\rho) A^{a}_\mu\nonumber \\
\left. +
g^{\rho\mu}(\partial^\nu A^{c*}_\rho) A^{a}_\mu A^{b*}_\nu - g^{\rho\mu} A^{c*}_\rho (\partial^\nu A^{a}_\mu) A^{b*}_\nu
\right\}
d^3x dt\,,
\gal
where $A^{a}_{\mu}$ is the potential of the incident gluon and $A^{b*}_\nu $, $A^{c*}_\rho$ are the potentials of the final gluons. The subscripts indicating their momenta and polarizations are omitted for brevity. Substituting \eq{bb4} allows representing \eq{dc5} in the following form:
\ball{dc7}
S(g\to gg)= i\frac{gf^{abc}}{V^{3/2}}\left\{ \b\epsilon^{*}_{\lambda'}(p')\cdot \b\epsilon_{\lambda_0}(p) \,\, (\b p+\b p')\cdot\b\epsilon^{c*}_\lambda(k)+ 
\b\epsilon^{*}_\lambda(k)\cdot \b\epsilon^{*}_{\lambda'}(p') \,\, (\b k-\b p')\cdot \b\epsilon_{\lambda_0}(p)\right. \nonumber\\
\left. - \b\epsilon^{*}_\lambda(k)\cdot \b\epsilon_{\lambda_0}(p)\,\, (\b k+\b p)\cdot \b\epsilon^{*}_{\lambda'}(p')\right\}
(2\pi)^3\delta(\b p+\b p'-\b k)G(\b p,\b k)\,,
\gal
where I employed a short-hand notation for the overlap function:
\ball{dc9}
G(\b p,\b k)= \int_{-\infty}^\infty a_p(t)a_k^*(t)a_{p'}^*(t)dt\,.
\gal
The three amplitudes $a(t)$ are given by \eq{ba13} with the subscript indicating the relevant gluon's momentum.  Noting that 
\ball{dc11}
p-p'-k= -\frac{k_\bot^2}{2p_zx(1-x)}\,,
\gal
yields: 
\bal
G(\b p,\b k)= \frac{1}{(2p_z)^{3/2}\sqrt{x(1-x)}}\int_{-\infty}^\infty  \exp\left\{ 
\frac{itk_\bot^2}{2p_zx(1-x)}+\frac{i}{2}(\lambda_0-\lambda'-\lambda)\int_0^tb_0(t')dt'- \epsilon t\sgn t
\right\}dt\label{dc13}\\
=\frac{2}{(2p_z)^{3/2}\sqrt{x(1-x)}}\int_0^\infty e^{-\epsilon t+ \frac{i}{2}(\lambda_0-\lambda'-\lambda)\int_0^t b_0''(t')dt'}\cos\left[- \frac{k_\bot^2}{2x(1-x)p_z}t-\frac{1}{2}(\lambda_0-\lambda'-\lambda)\int_0^t b_0'(t')dt'\right]dt\,.\label{dc14}
\gal

The gluon splitting probability is given by:
\ball{dc15}
&dw(g\to gg)= \frac{1}{2(N_c^2-1)}\sum_{\lambda_0,\lambda,\lambda'}\left|S(g\to gg)\right|^2\frac{Vd^3k}{(2\pi)^3}\frac{Vd^3p'}{(2\pi)^3}
= \frac{g^2N_c}{2}\sum_{\lambda_0,\lambda,\lambda'} |G(\b p,\b k)|^2 \frac{d^3k}{(2\pi)^3}\nonumber\\
&\times 
 \left| \b\epsilon^{*}_{\lambda'}(p')\cdot\b \epsilon_{\lambda_0}(p) \,\, (\b p+\b p')\cdot\b \epsilon^{*}_\lambda(k)+ 
\b\epsilon^{*}_\lambda(k)\cdot \b\epsilon^{*}_{\lambda'}(p') \,\, (\b k-\b p')\cdot\b \epsilon_{\lambda_0}(p) - \b\epsilon^{*}_\lambda(k)\cdot \b\epsilon_{\lambda_0}(p)\,\, (\b k+\b p)\cdot \b\epsilon^{*}_{\lambda'}(p')\right|^2\,.
\gal
The products of the polarization vectors can be computed using \eq{bb5} or utilizing the explicit forms \eq{dc3a}-\eq{dc3c}. Either way, the result is:
\ball{dc17}
dw(g\to gg)= \frac{g^2N_c}{2}\sum_{\lambda_0,\lambda,\lambda'} \frac{k_\bot^2}{2}
\left\{
-\frac{1+\lambda_0\lambda'}{x}+1-\lambda\lambda'-\frac{1+\lambda\lambda_0}{1-x}
\right\}^2 
|G(\b p,\b k)|^2 \frac{d^3k}{(2\pi)^3}\,.
\gal

\subsubsection{Constant $b_0$}

In the particular case of constant $b_0$, the overlap function $G$ is proportional to the delta function expressing energy conservation. Upon integrating over time in \eq{dc13}, the gluon production rate reads
\ball{dc21}
\frac{d\dot w(g\to gg)}{dx d^2k_\bot}= \frac{N_cg^2b_0}{2^6\pi^2}x(1-x)\sum_{\lambda_0,\lambda,\lambda'} (\lambda+\lambda'-\lambda_0)
\left\{
-\frac{1+\lambda_0\lambda'}{x}+1-\lambda\lambda'-\frac{1+\lambda\lambda_0}{1-x}
\right\}^2 \nonumber\\
\times 
\delta\left(k_\bot^2-x(1-x)(\lambda+\lambda'-\lambda_0)b_0p_z \right)\,.
\gal
After summing over the gluon polarizations, one derives:
\ball{dc24}
\frac{d\dot w(g\to gg)}{dx d^2k_\bot}= \frac{N_cg^2b_0}{8\pi^2}\left\{ \frac{x}{1-x}+\frac{1-x}{x}+x(1-x)\right\}
\delta\left(k_\bot^2-x(1-x)(\lambda+\lambda'-\lambda_0)b_0p_z \right)\,,
\gal
in agreement with the results of \cite{Hansen:2024rlj}. The expression in the braces, multiplied by $2N_c$, represents the nonsingular part of the splitting function $P_{gg}(x)$.

The delta function in \eq{dc24} is nonvanishing only when $(\lambda+\lambda'-\lambda_0)b_0$ is positive. Assume for definitiveness that $b_0>0$. Then, this condition requires that the sum of helicities of final gluons exceed the helicity of the initial gluon: $\lambda_0<\lambda+\lambda'$. This is only possible in the following four channels: 
\begin{enumerate}[label=(\roman*)]
\item $\lambda_0=1$, $\lambda=\lambda'=1$,
\item $\lambda_0=-1$, $\lambda=\lambda'=1$,
\item $\lambda_0=\lambda'=-1$, $\lambda=1$, 
\item $\lambda_0=\lambda=-1$, $\lambda'=1$.
\end{enumerate}
Out of the four channels, channel (ii) is suppressed at high energies, as evidenced by the vanishing of \eq{dc21}. 

The apparent asymmetry between the two gluon polarizations results in the polarization of the total gluon yield.

\subsubsection{$b_0(t)=A+ B\tanh\frac{t}{\tau}$}

\begin{figure}[ht]
\begin{tabular}{cc}
      \includegraphics[width=0.45\linewidth]{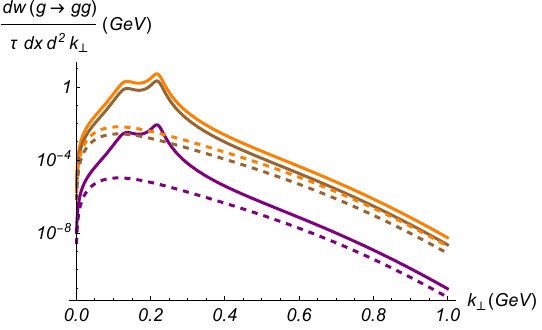} &
       \includegraphics[width=0.45\linewidth]{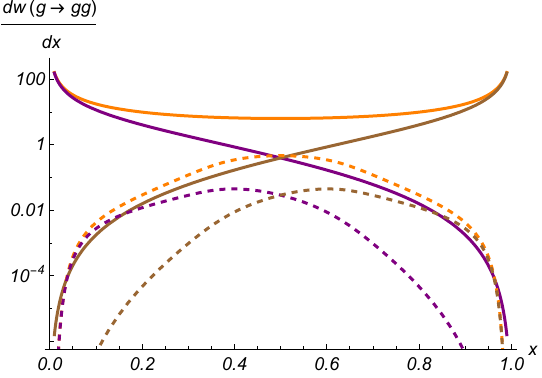} \
      \end{tabular}
  \caption{The gluon spectrum produced by the process $g(\lambda_0)\to g(\lambda)+g(\lambda')$. The orange lines represent $\lambda_0=1 , \lambda=1, \lambda'= 1$, the purple lines $\lambda_0=-1 , \lambda=1, \lambda'= -1$, and the brown lines  $\lambda_0=-1 , \lambda=-1, \lambda'= 1$. The solid lines indicate $\omega_p=0$, and the dashed line $\omega_p=0.3$~GeV.
The chiral magnetic conductivity is $b_0(t)=A_1+ B_1\tanh\frac{t}{\tau}$ with $A=10$~MeV, $B=-5$~MeV, and $\tau=25$~GeV$^{-1}\approx$ 5~fm/c.  The resonance width $\epsilon=1$~MeV. The incident gluon's energy is $E=20$~GeV. The produced gluons carry $x=0.8$ and $1-x=0.2$ of the incident gluon's longitudinal momentum, respectively.}\label{fig:ggg}
\end{figure}

Substituting $\Lambda= \Delta \lambda= \lambda_0-\lambda'-\lambda$ and $\alpha= p-p'-k$ in \eq{j5}, where the latter is given by \eq{dc11},  one can compute the overlap function \eq{dc14}:
\ball{ja21}
G(\b p, \b k) = \frac{\tau}{(2p_z)^{3/2}\sqrt{x(1-x)}}\frac{ 2^{-i\Delta\lambda B \tau/2-1}}{\Gamma(-i\Delta\lambda B \tau/2)}
\Gamma\left[\frac{i\tau}{2}\left(-\frac{\Delta\lambda }{2}(A+B)-\frac{k_\bot^2}{2x(1-x)p_z} -i\epsilon\right) \right]\nonumber\\
\times
\Gamma\left[\frac{i\tau}{2}\left(\frac{\Delta\lambda }{2}(A-B)+\frac{k_\bot^2}{2x(1-x)p_z} -i\epsilon\right) \right]\,.
\gal
The gluon spectrum \eq{dc17} now reads
\ball{dcj1}
\frac{dw(g\to gg)}{dx d^2k_\bot}&= \frac{g^2N_c}{2}\frac{p_z}{(2\pi)^3}\sum_{\lambda_0,\lambda,\lambda'} \frac{k_\bot^2}{2}
\left\{
-\frac{1+\lambda_0\lambda'}{x}+1-\lambda\lambda'-\frac{1+\lambda\lambda_0}{1-x}
\right\}^2 
\frac{\tau^2}{(2p_z)^{3}x(1-x)}\frac{ 1}{4\left|\Gamma(-i\Delta\lambda B \tau/2)\right|^2}\nonumber\\
&\times 
\left|\Gamma\left[\frac{i\tau}{2}\left(-\frac{\Delta\lambda }{2}(A+B)-\frac{k_\bot^2}{2x(1-x)p_z} -i\epsilon\right) \right]
\Gamma\left[\frac{i\tau}{2}\left(\frac{\Delta\lambda }{2}(A-B)+\frac{k_\bot^2}{2x(1-x)p_z} -i\epsilon\right) \right]\right|^2\,.
\gal
It is represented by solid lines in \fig{fig:ggg}.

The plasma oscillations are accounted for by replacing $k_\bot^2\to k_\bot^2 +\omega_p^2(x^2-x+1)$ in \eq{dc14} and in the arguments of gamma functions in \eq{dcj1}.  The result is shown as dashed lines in \fig{fig:ggg}. It is seen that the effect of finite $\omega_p$ is strongest in this gluon splitting channel. The is because the perturbative splitting amplitude is divergent at $x\to 0$ and $x\to 1$ as indicated by the solid lines in \fig{fig:ggg}. Since this region gives logarithmically enhanced contributions to the total multiplicity, and since finite $\omega_p$ cuts these regions off, the result is a relatively strong suppression of the total gluon multiplicity. 


\section{Energy Loss}\label{sec:m}

\begin{figure}[ht]
      \includegraphics[width=0.7\linewidth]{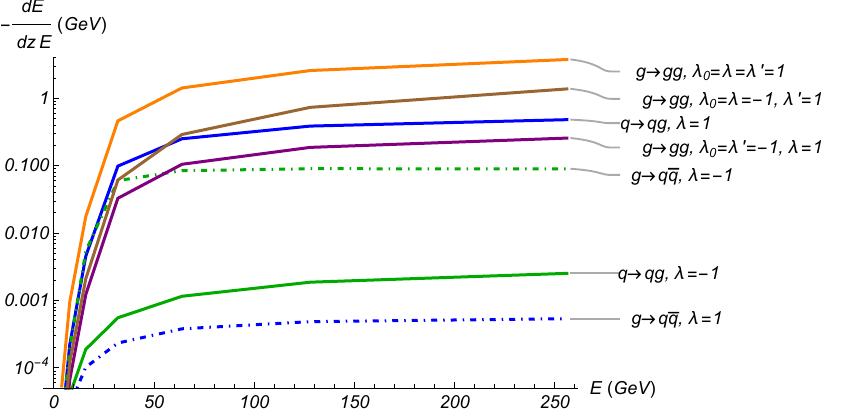} 
  \caption{The energy loss for different channels as a function of the incident parton energy $E$. The color coding is the same as in Figs.~\ref{fig:qqg},\ref{fig:gqq}, and \ref{fig:ggg}. Solid blue and green lines represent the $q\to q+g$ channel, and dash-dotted blue and green lines represent the $g\to q+\bar q$ channel.  The chiral magnetic conductivity is $b_0(t)=A_1+ B_1\tanh\frac{t}{\tau}$ with $A=10$~MeV, $B=-5$~MeV, and $\tau=25$~GeV$^{-1}\approx$ 5~fm/c.  The resonance width $\epsilon=1$~MeV. The quark's mass and the plasma frequency are set to be  $m=\omega_p=300$~MeV.}\label{fig:eloss}
\end{figure}

The energy loss per unit length experienced by an incident particle propagating through the chiral medium is given by
\ball{m1}
-\frac{dE(a\to bc)}{dz}= \int_0^E\omega \frac{dw(a\to bc)}{d\omega}d\omega = E\int_0^1x\frac{dw(a\to bc)}{dx}dx\,,
\gal
where $a$,$b$ and $c$ stand for $q$, $\bar q$ and $g$. The dependence of the energy loss on the incident particle energy is shown in \fig{fig:eloss}. The amount of the energy loss due to the chiral magnetic effect is comparable to conventional processes making it potentially significant for phenomenological implications. In particular, the gluon production channels exhibit a dominant role, which is logarithmically enhanced compared to the pair production. This occurs because the splitting functions are identical to those in conventional splittings, except for polarization dependence.

\section{Conclusions}\label{sec:s}

I computed the spectra of quarks, gluons and photons produced by ultrarelativistic particles in the presence of an arbitrary time-dependent chiral magnetic current. Eqs.~\eq{ca9},\eq{ca11}, and \eq{ca19} describe the photon and gluon spectra produced by an ultrarelativistic  quark in the chiral Cherenkov processes $q\to q+\gamma$ and $q\to q+g$. Eqs.~\eq{cb9},\eq{cb1},\eq{cb21} describe the quark spectrum produced by an ultrarelativistic gluon of photon in the pair production processes $\gamma \to q+\bar q$ and $g \to q+\bar q$. Eqs.~\eq{dc17},\eq{dc14} describe the gluon spectrum produced by an ultrarelativistic gluon in the process $g\to g+g$. These formulas only include contributions proportional to $b_0$, which arise due to the chiral magnetic effect, whereas the well-known conventional contributions are omitted.

Using a specific model $b_0(t)=A+ B\tanh\frac{t}{\tau}$ that describes the net chirality relaxation in the original $P$-odd domain, I derived the analytical expressions for the corresponding spectra given by \eq{caj3},\eq{cbj1} and \eq{dcj1}. The essential feature of this model is that it provides a smooth approximation of the step function. The details of the transition between the two limiting values $b_0\to \pm \infty$ are not essential. For practical illustration, the spectra \eq{caj3},\eq{cbj1} and \eq{dcj1} were applied to relativistic heavy-ion collisions, assuming an initial and final chiral magnetic conductivity of $15$~MeV, and 5~MeV, respectively, and a relaxation time of 5~fm/$c$. The results are presented in Figs.~\ref{fig:qqg},\ref{fig:gqq} and \ref{fig:ggg}. As expected for chiral Cherenkov radiation, the spectra exhibit a strong chirality dependence. For comparison, Figs.~\ref{fig:qqg-null},\ref{fig:gqq-null} and \ref{fig:ggg-null} in the Appendix show the spectra for an initial chiral magnetic conductivity of $15$~MeV that vanishes at late times. While the spectra in Figs.~\ref{fig:qqg}-\ref{fig:ggg} and Figs.~\ref{fig:qqg-null}-\ref{fig:ggg-null} are qualitatively similar, there are small quantitative differences. In any case, the uncertainty associated with the value of the plasma frequency has a larger effect than the uncertainty associated with the late-time value of $b_0$.

The spectra show a characteristic resonance structure that was noted in previous publications \cite{Tuchin:2025stl,Tuchin:2025bll}. The sensitivity of the spectra, in general, and this structure in particular, to collective plasma oscillations was studied by incorporating a finite plasma frequency $\omega_p$ into the gluon spectrum. As shown in Figs.~\ref{fig:qqg},\ref{fig:gqq} and \ref{fig:ggg}, the gluon spectrum is suppressed, while the quark spectrum is enhanced due to the dependence of the phase space volume on $\omega_p$. 

The amount of energy lost by a particle in the relaxing $P$-odd domain due to the chiral magnetic effect is shown in \fig{fig:eloss} and \fig{fig:eloss-null}  for each process. The strong variation of the energy loss with gluon polarization implies significant jet polarization in quark-gluon plasma. This observation may serve as an effective tool for studying parity-violating processes in hot nuclear matter.

The calculations in this work are performed within the Chern–Simons effective theory, where the chiral magnetic background is incorporated directly into the gauge-field sector. An important open question for future work is to clarify the relation between this formulation and the alternative fermionic axial-coupling approach not considered here.

\acknowledgments

This work  was supported in part by the U.S. Department of Energy under Grant No.\ DE-SC0023692.

\appendix
\section{Parton spectra for $b_0(\infty)=0$}\label{appA}

\begin{figure}[ht]
\begin{tabular}{cc}
      \includegraphics[width=0.45\linewidth]{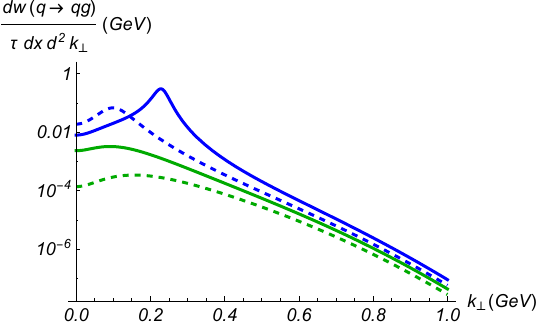} &
       \includegraphics[width=0.45\linewidth]{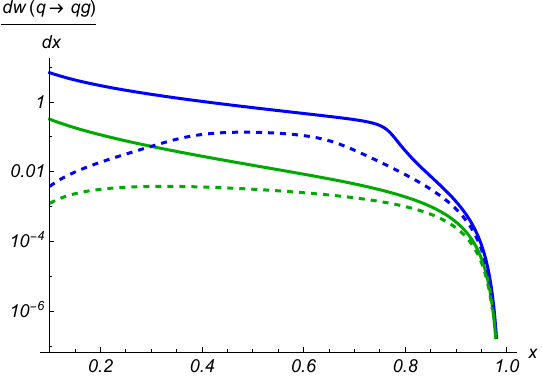} \
      \end{tabular}
  \caption{Same as in \fig{fig:qqg}, except that $A=-B=7.5$~MeV.}\label{fig:qqg-null}
\end{figure}

\begin{figure}[ht]
\begin{tabular}{cc}
      \includegraphics[width=0.45\linewidth]{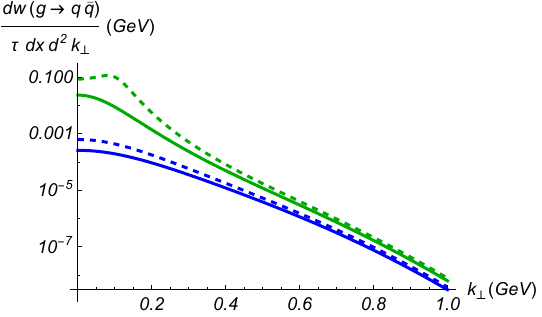} &
       \includegraphics[width=0.45\linewidth]{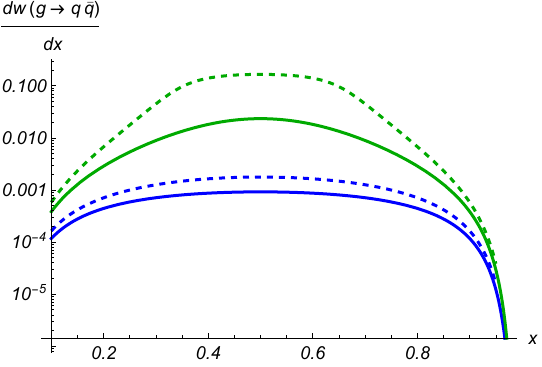} \
      \end{tabular}
  \caption{Same as in \fig{fig:gqq}, except that $A=-B=7.5$~MeV.}\label{fig:gqq-null}
\end{figure}

\begin{figure}[ht]
\begin{tabular}{cc}
      \includegraphics[width=0.45\linewidth]{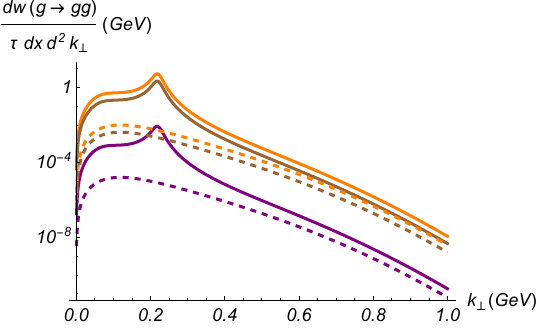} &
       \includegraphics[width=0.45\linewidth]{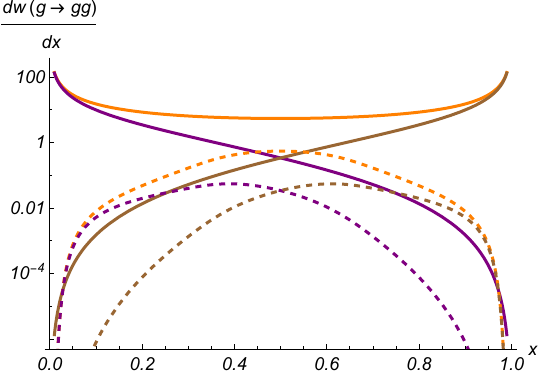} \
      \end{tabular}
  \caption{Same as in \fig{fig:ggg}, except that $A=-B=7.5$~MeV.}\label{fig:ggg-null}
\end{figure}

\begin{figure}[ht]
      \includegraphics[width=0.7\linewidth]{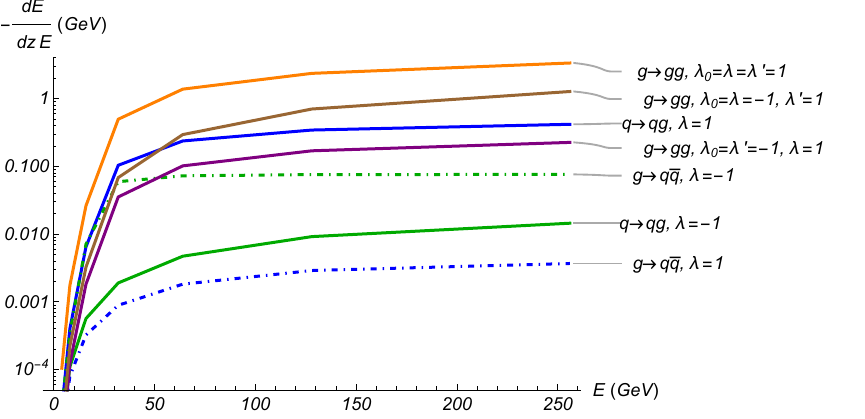} 
  \caption{Same as in \fig{fig:eloss}, except that $A=-B=7.5$~MeV.}\label{fig:eloss-null}
\end{figure}

\bibliography{anom-biblio}

\end{document}